\documentclass[twocolumn,showpacs,preprintnumbers]{revtex4}
\usepackage{amssymb}
\usepackage{amsmath}
\usepackage[english]{babel}
\usepackage[dvips]{graphicx}

\newcommand{\be}{\begin{equation}}
\newcommand{\ee}{\end{equation}}
\newcommand{\bea}{\begin{eqnarray}}
\newcommand{\eea}{\end{eqnarray}}

\begin{document}

\title{Wave scattering on a domain wall in a chain of $\mathcal{PT}$%
-symmetric couplers}
\author{Sergey V. Suchkov$^{1}$, Sergey V. Dmitriev$^{1}$, Boris A. Malomed$%
^{2,3}$, and Yuri S. Kivshar$^{4}$}
\affiliation{$^1$Institute for Metals Superplasticity Problems, Russian Academy of
Science, Ufa 450001, Russia \\
$^{2}$Department of Physical Electronics, School of Electrical Engineering,
Faculty of Engineering, Tel Aviv University, Tel Aviv 69978, Israel \\
$^{3}$ICFO-Institut de Ciencies Fotoniques, Mediterranean Technology Park,
Castelldefels (Barselona) 08860, Spain\thanks{%
Sabbatical address}\\
$^{4}$Nonlinear Physics Center, Research School of Physics and Engineering,
Australian National University, Canberra ACT 0200, Australia}

\begin{abstract}
We study wave propagation in linear arrays composed of pairs of conjugate
waveguides with balanced gain and loss, i.e. arrays of the $\mathcal{PT}$%
-symmetric couplers, where the linear spectrum is known to feature
high-frequency and low-frequency branches. We introduce a domain wall by
switching the gain and loss in a half of the array, and analyze the
scattering of linear waves on this defect. The analysis reveals two major
effects: amplification of both reflected and transmitted waves, and
excitation of the reflected and transmitted low-frequency and high-frequency
waves by the incident high-frequency and low-frequency waves, respectively.
\end{abstract}

\maketitle

\section{Introduction}

It is well established that non-Hermitian Hamiltonians, subject to the
constraint of the parity-time ($\mathcal{PT}$) symmetry, i.e., the
equilibrium between spatially separated loss and gain, which are set as
mirror images of each other, may give rise to entirely real spectra,
provided that the strength of the gain and loss does not exceeds a critical
level \cite{Bender}. Although, generally speaking, $\mathcal{PT}$-symmetric
settings belong to the class of dissipative systems, they can support
continuous families of both linear and nonlinear modes, thus resembling the
main properties of conservative systems. In fact, $\mathcal{PT}$-symmetric
models lie at the borderline between conservative and truly dissipative
dynamical systems.

It is straightforward to implement $\mathcal{PT}$-symmetric complex
potentials, $V(x)$, which must be subject to the aforementioned equilibrium
condition, $V(x)=V^{\ast }(-x)$, by symmetrically juxtaposing elements
accounting for the gain and loss~\cite{Ruschhaupt}. This possibility was
elaborated in a number of theoretical~\cite{theory}-\cite{we} and
experimental~\cite{experiment} studies.

In optics, the basic $\mathcal{PT}$-symmetric element can be realized as a
pair of linearly coupled waveguides, one with a lossy core and the other one
carrying a matched compensating gain~\cite{Coupler}. A chain composed of
such coupled elements was introduced in Ref.~\cite{OL}, assuming that each
amplified and dissipative waveguide was linearly linked to an adjacent
waveguide of the opposite sign, belonging to the neighboring pair.

Another type of such a $\mathcal{PT}$-symmetric system was proposed in Ref.~%
\cite{we}, with the gain- and loss-carrying units coupled to their
neighboring counterparts of the same sign [see Fig.~\ref{Fig1}(a) below].
Assuming that each unit also carried the conservative cubic nonlinearity,
discrete solitons were found in this setting.

The subject of this paper is a system of the general same type (although
without nonlinearity), but including a defect in the form of the domain wall
(DW), as shown below in Fig.~\ref{Fig1}(b). A natural dynamical problem to
consider in such an array, in addition to the DW itself, is the scattering
of linear waves on the DW, which is another subject of the present work (the
scattering of linear waves on an isolated $\mathcal{PT}$-symmetric complex,
including such specific features as amplification of transmitted and
reflected waves and Fano resonances, was analyzed in Refs. \cite%
{Miroshnichenko} and \cite{DS}). Below we demonstrate that the scattering of
waves on the DW gives rise to nontrivial effects, including the
transformations between different branches of the traveling-wave modes and
their amplification. The fact that we consider the scattering of incident
waves with real frequencies, and the generation of transmitted and reflected
waves which are carried by real frequencies too, implies that we are dealing
with the case when the spectrum of the $\mathcal{PT}$-invariant system is
purely real, under the condition that the gain-loss coefficient is kept
below the critical value [see Eq. (\ref{restriction}) below].

It is relevant to mention that this scattering problem is related to the
analysis of transport and scattering processes governed by non-Hermitian
Hamiltonians \cite{transport}. In the general case, such Hamiltonians are
not subject to the $\mathcal{PT}$-symmetry constraint, therefore the
corresponding spectrum is complex. In particular, the analysis of the
generic spectra demonstrates that they contain a few eigenvalues with
especially large imaginary parts. The rapid decay of the corresponding
eigenstates may be considered as an analog of superradiance in optics \cite%
{superradiance}, and, naturally, it strongly affects dynamical
features of such systems. It remains to understand if
$\mathcal{PT}$-symmetric systems may give rise to similar
``superradiant" states in the respective complex spectrum (above the
transition from the real spectrum).

The paper is organized as follows. Section~\ref{Model} introduces the model,
and also it discusses the geometry of a domain-wall defect introduced into
the chain. Analytical results for the transmission and reflection
coefficients are presented in Sec.~\ref{AnalyticalResults}, whereas the
dependence of the scattering coefficients on the system parameters is
analyzed in Sec.~\ref{Analysis}. Finally, Sec.~\ref{Conclusions} concludes
the paper.

\section{The model}

\label{Model}

We consider the array of paired waveguides with balanced gain and loss ($%
\gamma $), i.e., $\mathcal{PT}$-symmetric couplers, as shown in Fig.~\ref%
{Fig1}(a). The linear-coupling constant for the waveguides in the $\mathcal{%
PT}$ elements is normalized to be $1$, while the coefficient of the linear
coupling between adjacent elements in the array is $C_{1}$. Thus, the model
is based on the following system of linear Schr\"{o}dinger equations:
\begin{eqnarray}
\frac{d}{dz}\left\{
\begin{array}{c}
u_{n} \\
v_{n}%
\end{array}%
\right\} &=&\gamma \left\{
\begin{array}{c}
u_{n} \\
-v_{n}%
\end{array}%
\right\} +i\left\{
\begin{array}{c}
v_{n} \\
u_{n}%
\end{array}%
\right\}  \notag \\
&&+iC_{1}\left\{
\begin{array}{c}
u_{n+1}+u_{n-1}-2u_{n} \\
v_{n+1}+v_{n-1}-2v_{n}%
\end{array}%
\right\} ,  \label{lpwd}
\end{eqnarray}%
where $u_{n}(z)$ and $v_{n}(z)$ are the complex amplitudes in the amplified
and damped waveguides at each site of the array. Actually, Eq. (\ref{lpwd})
is a linearized version of the model introduced in recent work \cite{we}.

\begin{figure}[tbp]
\includegraphics[width=\columnwidth]{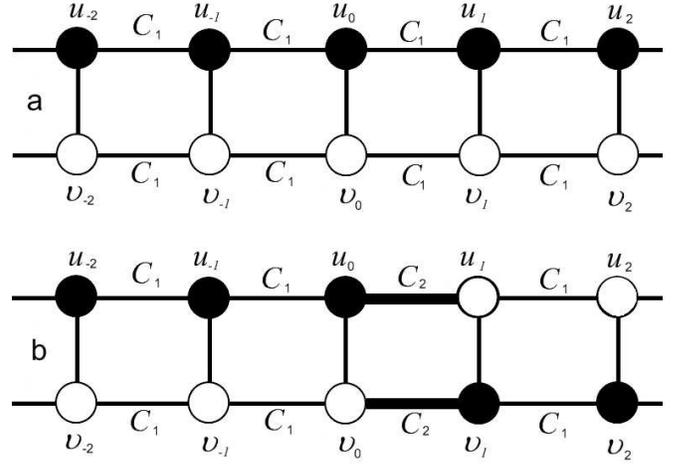}
\caption{(a) A schematic of the chain of $\mathcal{PT}$-symmetric couplers
composed of waveguides with gain and loss, which are designated by dark and
bright circles, respectively. (b) The model with the domain wall, created by
inverting gain and loss in a half of the chain.}
\label{Fig1}
\end{figure}

The DW in the array is created by switching the gain and loss in a half of
the chain; generally, the constant of the linear coupling between the
halves, $C_{2}$, may be different from the regular value, $C_{1}$ [see Fig.~%
\ref{Fig1}(b)]. Thus, the array with the embedded DW is described by the
following equations:
\begin{eqnarray}
\frac{du_{n}}{dz} &=&\gamma u_{n}+iv_{n}+iC_{1}(u_{n+1}+u_{n-1}-2u_{n}),
\notag \\
\frac{dv_{n}}{dz} &=&-\gamma v_{n}+iu_{n}+iC_{1}(v_{n+1}+v_{n-1}-2v_{n}),
\notag \\
n &\leq &-1,  \label{vnormleft}
\end{eqnarray}
\begin{eqnarray}
\frac{du_{n}}{dz} &=&-\gamma u_{n}+iv_{n}+iC_{1}(u_{n+1}+u_{n-1}-2u_{n}),
\notag \\
\frac{dv_{n}}{dz} &=&\gamma v_{n}+iu_{n}+iC_{1}(v_{n+1}+v_{n-1}-2v_{n}),
\notag \\
n &\geq &2,  \label{vnormright}
\end{eqnarray}%
\begin{eqnarray}
\frac{du_{0}}{dz} &=&\gamma
u_{0}+iv_{0}+iC_{1}(u_{-1}-u_{0})+iC_{2}(u_{1}-u_{0}),  \notag
\label{vnormcentr0} \\
\frac{dv_{0}}{dz} &=&-\gamma
v_{0}+iu_{0}+iC_{1}(v_{-1}-v_{0})+iC_{2}(v_{1}-v_{0}),\,\,\,\,\,\, \\
\frac{du_{1}}{dz} &=&-\gamma
u_{1}+iv_{1}+iC_{1}(u_{2}-u_{1})+iC_{2}(u_{0}-u_{1}),  \notag \\
\frac{dv_{1}}{dz} &=&\gamma
v_{1}+iu_{1}+iC_{1}(v_{2}-v_{1})+iC_{2}(v_{0}-v_{1}),  \label{vnormcentr1}
\end{eqnarray}%

\section{Analytical results}

\label{AnalyticalResults}

First, we consider the wave propagation in the array without the DW. The
corresponding solution to Eq.~(\ref{lpwd}) in looked for as
\begin{equation}
\left\{
\begin{array}{c}
u_{n} \\
v_{n}%
\end{array}%
\right\} =\left\{
\begin{array}{c}
e^{i\delta } \\
1%
\end{array}%
\right\} \exp [i(kn-\omega z)],  \label{genform}
\end{equation}%
where wavenumber $k$ may be complex, while frequency $\omega $ is real.
Substituting Eq.~(\ref{genform}) into Eq.~(\ref{lpwd}), one finds the
high-frequency (HF) and low-frequency (LF) branches of the dispersion
relation for the linear waves, denoted by subscripts $h$ and $l$,
respectively:
\begin{eqnarray}
\omega _{h} &=&2C_{1}\left( 1-\frac{e^{ik}+e^{-ik}}{2}\right) -\cos \delta
_{h},  \notag \\
\sin \delta _{h} &=&-\gamma ,\quad \cos \delta _{h}=-\sqrt{1-\gamma ^{2}},
\label{cond_h} \\
\omega _{l} &=&2C_{1}\left( 1-\frac{e^{ik}+e^{-ik}}{2}\right) -\cos \delta
_{l},  \notag \\
\sin \delta _{l} &=&-\gamma ,\quad \cos \delta _{l}=\sqrt{1-\gamma ^{2}}.
\label{cond_l}
\end{eqnarray}%
In particular, for $k=i\kappa $ with real $\kappa $ we obtain an
exponential-wave (EW) solution to Eq.~(\ref{lpwd}), while $k=i\kappa +\pi $
gives rise to a staggered exponential-wave (SEW) one. Continuous-wave (CW)
solutions correspond to $k=\kappa $, i.e., real wavenumbers. Only these
three types of the waves admit real frequencies $\omega $, which, in
addition, requires that the gain/loss coefficient must be smaller than the
constant of the coupling between the amplified and dissipative waveguides in
each $\mathcal{PT}$\ element, i.e.,
\begin{equation}
|\gamma |\leq 1.  \label{restriction}
\end{equation}

\begin{figure}[tbp]
\includegraphics[width=\columnwidth]{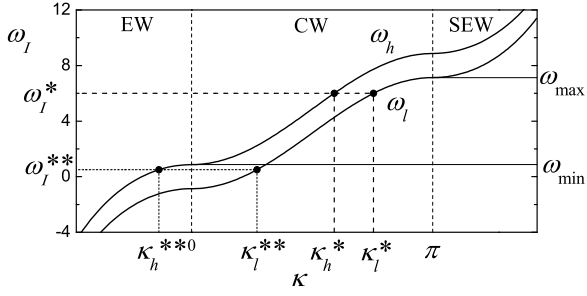}
\caption{Spectrum of the array without the domain wall, obtained from Eq.~(%
\protect\ref{lpwd}) for $C_{1}=2$, $\protect\gamma =0.5$. The HF and LF
branches, $\protect\omega _{h}$ and $\protect\omega _{l}$, are defined by
Eqs.~(\protect\ref{cond_h}),(\protect\ref{cond_l}), with $k=i\protect\kappa $
for $-\infty <\protect\kappa <0$ (the EW region), $k=\protect\kappa $ for $%
0\leq \protect\kappa \leq \protect\pi $ (the CW region), and $k=\protect\pi %
+i(\protect\kappa -\protect\pi )$ for $\protect\pi <\protect\kappa <\infty $
(the SEW region).}
\label{Fig2}
\end{figure}

An example of the spectrum determined by Eqs. (\ref{cond_h}) and (\ref%
{cond_l}) is presented in Fig.~\ref{Fig2}. In the EW region, $-\infty
<\kappa <0$, $\omega _{h}(i\kappa )$ and $\omega _{l}(i\kappa )$ are
plotted. The range of $0\leq \kappa \leq \pi $ corresponds to the CW
solutions, where we show the frequencies as $\omega _{h}(\kappa )$ and $%
\omega _{l}(\kappa )$. At $\kappa >\pi $, we have the SEW region, in which
the branches of the dispersion relation are plotted as $\omega _{h}(\pi
+i[\kappa -\pi ])$ and $\omega _{l}(\pi +i[\kappa -\pi ])$.

Now we proceed to the scattering of linear waves on the DW, within the
framework of Eqs. (\ref{vnormleft})-(\ref{vnormcentr1}). To this end, we
consider an incident wave, with frequency $\omega _{I}$ (and the intensity
set equal to $1$), which approaches the DW from the left. Only the case when
the incident wave belongs to the LF branch is studied below in an explicit
form, as the case of the HF incident wave can be considered similarly.

We look for the scattering solution to Eqs.~(\ref{vnormleft})-(\ref%
{vnormcentr1}) as follows:
\begin{eqnarray}  \label{lpch}
&&\left\{
\begin{array}{c}
u_{n} \\
v_{n}%
\end{array}%
\right\}=e^{i(k_{l}n-\omega _{I}z)}\left\{
\begin{array}{c}
e^{i\delta _{l}} \\
1%
\end{array}%
\right\} \\
&&+R_{h}e^{i(-k_{h}n-\omega _{I}z)}\left\{
\begin{array}{c}
e^{i\delta _{h}} \\
1%
\end{array}%
\right\} +R_{l}e^{i(-k_{l}n-\omega _{I}z)}\left\{
\begin{array}{c}
e^{i\delta _{l}} \\
1%
\end{array}%
\right\},  \notag
\end{eqnarray}
for $n\leq 0$, and
\begin{eqnarray}
\left\{
\begin{array}{c}
u_{n} \\
v_{n}%
\end{array}%
\right\} &=&T_{h}e^{i(k_{h}n-\omega _{I}z)}\left\{
\begin{array}{c}
1 \\
e^{i\delta _{h}}%
\end{array}%
\right\}  \notag \\
&&+T_{l}e^{i(k_{l}n-\omega _{I}z)}\left\{
\begin{array}{c}
1 \\
e^{i\delta _{l}}%
\end{array}%
\right\} ,  \label{rpch}
\end{eqnarray}%
for $n\geq 1$. Here the characteristic of the plane-wave components of the
solution are $\sin \delta _{l}=\sin \delta _{h}=-\gamma $, $\cos \delta
_{l}=-\cos \delta _{h}=\sqrt{1-\gamma ^{2}}$, and $k_{h}$, $k_{l}$ are to be
found from Eqs. (\ref{cond_h}) and Eqs. (\ref{cond_l}), respectively, where
we set $\omega _{h}=\omega _{l}=\omega _{I}$.

Amplitudes $R_{l},\,T_{l}$ and $R_{h},\,T_{h}$, defined in expressions (\ref%
{lpch}) and (\ref{rpch}) are complex reflection and transmission
coefficients for the LF and HF waves. Substituting these expressions into
Eqs. (\ref{vnormleft})-(\ref{vnormcentr1}), we obtain
\begin{eqnarray}
R_{l} &=&\frac{e^{ik_{l}}}{D}\biggl\{e_{2}\,(e_{3}-1)+(1-\gamma ^{2})\Bigl[%
e_{1}-2e^{2ik_{l}}  \notag \\
&+&e_{3}(e_{1}+4(e^{ik_{h}}-1)-2e_{3})-\overline{C}^{2}\left(
1-e_{1}+e_{3}\right) ^{2}  \notag \\
&+&\overline{C}\left( 1-e^{ik_{l}}\right) ^{2}\left( 1-e^{ik_{h}}\right)
\left( 1-3e^{ik_{h}}\right) \Bigr]\biggr\},  \notag \\
R_{h} &=&\frac{e^{ik_{h}}}{D}\left( 1-e^{2ik_{l}}\right) \,e_{2}\,\gamma
\,(\gamma +i\sqrt{1-\gamma ^{2}}),  \notag \\
T_{l} &=&\frac{1}{D}\left( -1+e^{2ik_{l}}\right) \left( -1+e^{ik_{h}}\right)
\times  \notag \\
&\,&\left[ -2e^{ik_{h}}+\overline{C}\left( -1+e^{ik_{h}}\right) \right]
\sqrt{1-{\gamma }^{2}},  \notag \\
T_{h} &=&-\frac{1}{D}\left( e^{2ik_{l}}-1\right) \left( e_{1}-2e_{3}+{%
\overline{C}}(e_{3}-e_{1}+1)\right) \times  \notag \\
&\,&\gamma \sqrt{1-\gamma ^{2}}\left( \gamma +i\,\sqrt{1-{\gamma }^{2}}%
\right)  \label{coeffwave}
\end{eqnarray}%
where we define%
\begin{eqnarray}
D \equiv {e_{2}}^{2}-\left( 1-{\gamma }^{2}\right) \biggl\{{e_{1}}%
^{2}-4e_{3}(e_{1}-e_{3})  \notag \\
+{\overline{C}}(e_{3}-e_{1}+1)\left[ 2e_{1}-4e_{3}+{\overline{C}}%
(e_{3}-e_{1}+1)\right] \biggr\},  \label{coeff1}
\end{eqnarray}%
\begin{eqnarray}
\overline{C} &\equiv &C_{1}/C_{2},\quad e_{1}\equiv e^{ik_{h}}+e^{ik_{l}}, \\
e_{2} &\equiv &e^{ik_{l}}-e^{ik_{h}},\quad e_{3}\equiv e^{i\,(k_{h}+k_{l})}.
\end{eqnarray}

The so obtained reflection and transmission coefficients can be analyzed for
the LF incident wave of any of the three types, EW, CW, or SEW. As seen from
Eqs. (\ref{lpch}) and (\ref{rpch}), the incident wave of the LF type
generates, generally speaking, both LF and HF reflected and transmitted wave
components, whose frequencies are identical to $\omega _{I}$.

We start by considering two examples, which are designated in Fig.~\ref{Fig2}%
. Taking $\omega _{I}=\omega _{I}^{\ast }$, we conclude that both $%
k_{h}=\kappa _{h}^{\ast }$ and $k_{l}=\kappa _{l}^{\ast }$ are in the CW
region, meaning that the reflected and transmitted HF and LF waves are all
of CW type (here, the asterisk \emph{does not} stand for the complex
conjugate). On the other hand, for $\omega _{I}=\omega _{I}^{\ast \ast }$ we
have $k_{h}=-i\kappa _{h}^{\ast \ast }$ and $k_{l}=\kappa _{l}^{\ast \ast }$%
, meaning that the incident wave is the LF of the CW type, while the
reflected and transmitted ones have the HF and LF components of the EW and
CW types, respectively.

Care should be taken as concerns the choice of the sign in front of
imaginary part of $k_{h}$ and $k_{l}$ for the EW. For the plus (minus) sign,
the reflected and transmitted waves of the EW type exponentially decrease
(increase) with the increase of the distance from the DW. Both cases are
physically meaningful in the $\mathcal{PT}$ system, but in the following we
focus only on the evanescent (exponentially decaying) EWs.

\begin{figure}[tbp]
\includegraphics[width=\columnwidth]{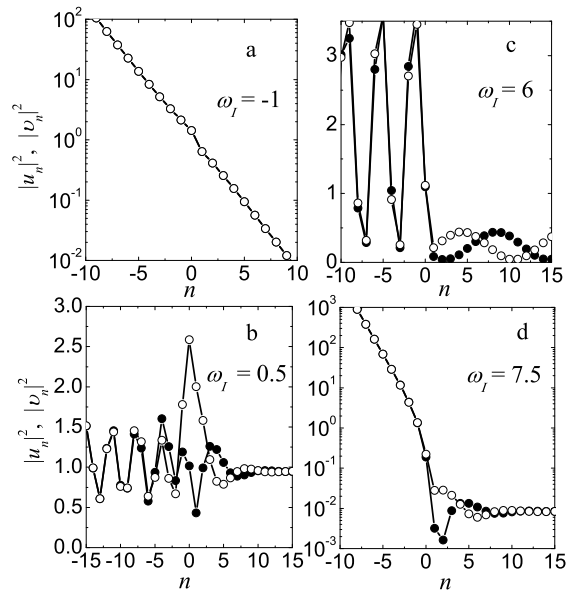}
\caption{Examples of solutions of the scattering problem in the array of
couplers for $C_{1}=2$, $C_{2}=1$, $\protect\gamma =0.5$, and different
values of the incident-wave's frequency, $\protect\omega _{I}$. (a) The
incident low-frequency exponential wave excites the high-frequency
evanescent wave, with $\protect\omega _{I}=-1$, $k_{h}=0.931843\,i$, $%
k_{l}=0.258102\,i$. (b) The incident low-frequency CW excites a
high-frequency exponential wave, with $\protect\omega _{I}=0.5$, $%
k_{h}=0.424603\,i$, $k_{l}=0.851981$. (c) The incident low-frequency CW
excites a high-frequency CW, with $\protect\omega _{I}=6$, $k_{h}=1.85823$, $%
k_{l}=2.36958$. (d) The incident low-frequency staggered exponential wave
excites a high-frequency CW, with $\protect\omega _{I}=7.5$, $k_{h}=2.28961$%
, $k_{l}=\protect\pi +0.424603\,i$. Dots and open circles show $|u_{n}|^{2}$
and $|v_{n}|^{2}$, respectively.}
\label{Fig3}
\end{figure}

Several generic examples of the wave-intensity profiles are present in Fig.~%
\ref{Fig3} in the array of couplers for different frequencies of the
incident wave, $\omega _{I}$, which are chosen with regard to the dispersion
relation displayed in Fig.~\ref{Fig2}. As mentioned above, the LF incident
wave generates reflected and transmitted waves belonging to the HF and LF
branches, with the same frequency $\omega _{I}$. The following four cases
are present in Fig.~\ref{Fig3}: (a) $\omega _{I}=-1$, with $%
k_{h}=0.931843\,i $, $k_{l}=0.258102\,i$, hence the incident wave is the EW
of the LF type, the corresponding HF and LF components also being EWs [see
Fig.~\ref{Fig3} (a), where the logarithmic scale is used on the vertical
axis]. The intensity of the incident wave increases with distance form the
DW, while the reflected and transmitted waves are evanescent. (b) $\omega
_{I}=0.5$, with $k_{h}=0.424603\,i$, $k_{l}=0.851981$, hence the incident LF
wave is a CW, while the reflected and transmitted waves include LF CW and HF
evanescent component, the latter rapidly decaying with the distance from the
defect [Fig.~\ref{Fig3}(b)]. (c) $\omega _{I}=6$, with $k_{h}=1.85823$, $%
k_{l}=2.36958$, in which case all the waves are of the CW type [Fig.~\ref%
{Fig3} (c)]. (d) $\omega _{I}=7.5$, with $k_{h}=2.28961$, $k_{l}=\pi
+0.424603\,i$, which makes the incident LF\ wave an SEW, while the reflected
and transmitted waves include the HF CW and LF SEW terms [see Fig.~\ref{Fig3}
(d), where the logarithmic scale is again used on the vertical axis].

Note that, at
\begin{equation}
\sqrt{1-\gamma ^{2}}>2C_{1},  \label{cross}
\end{equation}%
the splitting between the LF and HF branches in Fig.~\ref{Fig2} becomes so
large that their CW regions do not have common frequencies. Under this
condition, the case presented in Fig.~\ref{Fig3}(c), with all the wave
components being of the CW type, is impossible. In terms of $\omega _{\min }=%
\sqrt{1-\gamma ^{2}}$ and $\omega _{\max }=4C_{1}-\sqrt{1-\gamma
^{2}}$, which are the smallest and the largest frequencies of the
HF and LF branches in the CW region (see \ref{Fig2}),
condition~(\ref{cross}) is tantamount to $\omega _{\max }<\omega
_{\min }$.

In the following we take $C_{1}\geq 0.5$, which rules out condition (\ref%
{cross}) for all $0\leq \gamma <1$. We thus always include the possibility
of having all the reflected and transmitted waves of the CW type, excited by
the incident LF CW.

In the following Section we present the analysis of the transmission and
reflection coefficients as functions of parameters of the system, assuming,
as said above, $\omega _{\min }<\omega _{\max }$, and focusing on the range
of the incident-wave's frequency $\omega _{\min }<\omega _{I}<\omega _{\max
} $, when this wave is of the most relevant CW type. Further, it is
convenient to define the normalized frequency,
\begin{equation}
\omega _{I}^{\prime }=\frac{\omega _{I}-\omega _{\min }}{\omega _{\max
}-\omega _{\min },}
\end{equation}%
so that $\omega _{I}^{\prime }=0$ at $\omega _{I}=\omega _{\min }$, and $%
\omega _{I}^{\prime }=1$ at $\omega _{I}=\omega _{\max }$.

\section{Transmission and reflection coefficients}

\label{Analysis}

Here we analyze the transmission and reflection coefficients for the LF
incident wave, given by Eq.~(\ref{coeffwave}) and Eq.~(\ref{coeff1}),
varying parameters $\gamma $, $C_{1}$, and $\overline{C}\equiv C_{1}/C_{2}$.
As mentioned above, we consider only the case of the CW incident wave as the
most natural one. Then, two possibility may be expected, with the incident
LF wave exciting either HF-EW or HF-CW.

\begin{figure}[tbp]
\includegraphics[width=\columnwidth]{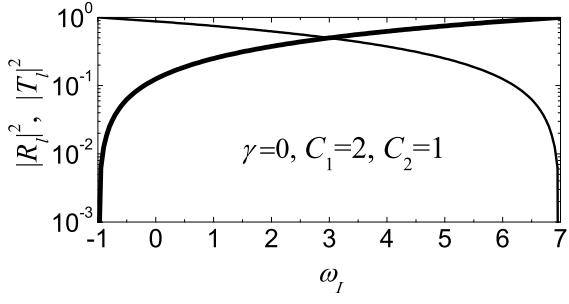}
\caption{$|R_{l}|^{2}$ (the thick line) and $|T_{l}|^{2}$ (the thin line) as
functions of $\protect\omega _{I}$ in the case of the incident low-frequency
CW for $C_{1}=2$, $C_{2}=1$, and $\protect\gamma =0$ (no gain and loss). In
this case, the high-frequency waves are not excited by the low-frequency
input ($R_{h}=0$, $T_{h}=0$).}
\label{Fig4}
\end{figure}

It is useful to study first the case with no gain and loss, $\gamma =0$.
Under this condition, Eq.~(\ref{coeffwave}) reduces to
\begin{eqnarray}
R_{l} &=&\frac{(\overline{C}-1)e^{ik_{l}}(e^{ik_{l}}-1)}{\overline{C}%
(e^{ik_{l}}-1)-2e^{ik_{l}}},  \notag \\
R_{h} &=&0,  \notag \\
T_{l} &=&-\frac{1+e^{ik_{l}}}{\overline{C}(e^{ik_{l}}-1)-2e^{ik_{l}}},
\notag \\
T_{h} &=&0.  \label{0}
\end{eqnarray}%
From here, we immediately arrive at the conclusion that the LF incident wave
excites only the LF reflected and transmitted waves, satisfying condition $%
|R_{l}|^{2}+|T_{l}|^{2}=1$ due to the energy conservation. Thus, the
excitation of the HF reflected and transmitted waves by the LF input is a
specific effect of the $\mathcal{PT}$ system, due to the presence of the
gain and loss in it. Expressions for $R_{l}$ and $T_{l}$\ in Eqs. (\ref{0})
become singular, with a vanishing denominator, only when $\overline{C}=1$
and $k_{l}=\pi $, which means the absence of the DW in the system (in which
case $T_{l}=1$ and $R_{l}=0$). Figure~\ref{Fig4} displays $|R_{l}|^{2}$ and $%
|T_{l}|^{2}$ as functions of $\omega _{I}$ for parameters $\gamma =0$ and $%
\overline{C}=2$.

In the presence of the gain and loss ($\gamma >0$) the solution of the
scattering problem for the LF incident wave contains $R_{h},T_{h}\neq 0$,
i.e., the HF components are excited. Note, however, that $R_{h}=T_{h}=T_{l}$
at $k_{l}=0$ or $k_{l}=\pi $, i.e., at the borders of the considered range
of the incident-wave's frequency, $\omega _{I}$. Also, $T_{l}=0$ if $k_{h}=0$
or $\overline{C}=2/(1-e^{-ik_{h}})$ (the latter is possible in the SEW
region, which is not under consideration here).

We now fix $\gamma =0.5$, $C_{1}=2$, and plot the (typical) results,
produced by Eqs. (\ref{coeffwave}) for $\overline{C}=2$, $\overline{C}=1$,
and $\overline{C}=0.5$.

\begin{figure}[tbp]
\includegraphics[width=\columnwidth]{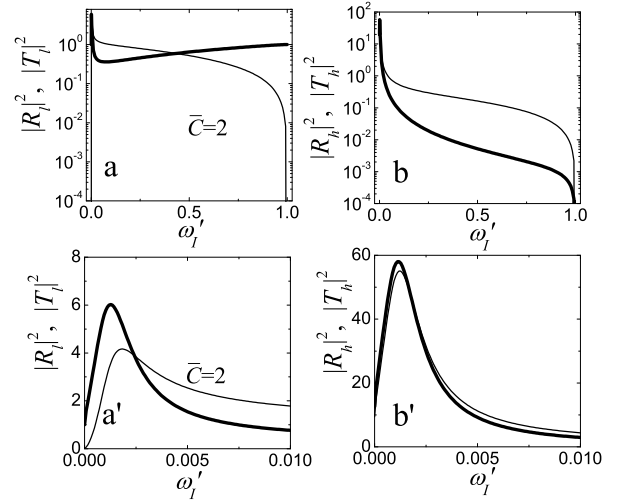}
\caption{Relative intensities of the reflected and transmitted waves as
functions of $\protect\omega _{I}^{\prime }$ in the case of the
low-frequency incident CW, for $\protect\gamma =0.5$ an $\overline{C}=2$.
Thick and thin lines depict $|R_{l}|^{2}$, $|R_{h}|^{2}$ in (a) and $%
|T_{l}|^{2}$, $|T_{h}|^{2}$ in (b), respectively. (a') and (b') are the
blowups of (a) and (b), respectively.}
\label{Fig5}
\end{figure}

Figure~\ref{Fig5} shows that, at $\overline{C}=2$, all four reflection and
transmission coefficients have finite maxima close to the point $\omega
_{I}^{\prime }=0.001$ ($\omega _{I}=0.872$). Note that (a') and (b') show
blowups of (a) and (b), respectively, in the vicinity of the maxima. Also, $%
T_{l}$ vanishes at $\omega _{I}^{\prime }=0$ ($\omega _{I}=0.866025$).

\begin{figure}[tbp]
\includegraphics[width=\columnwidth]{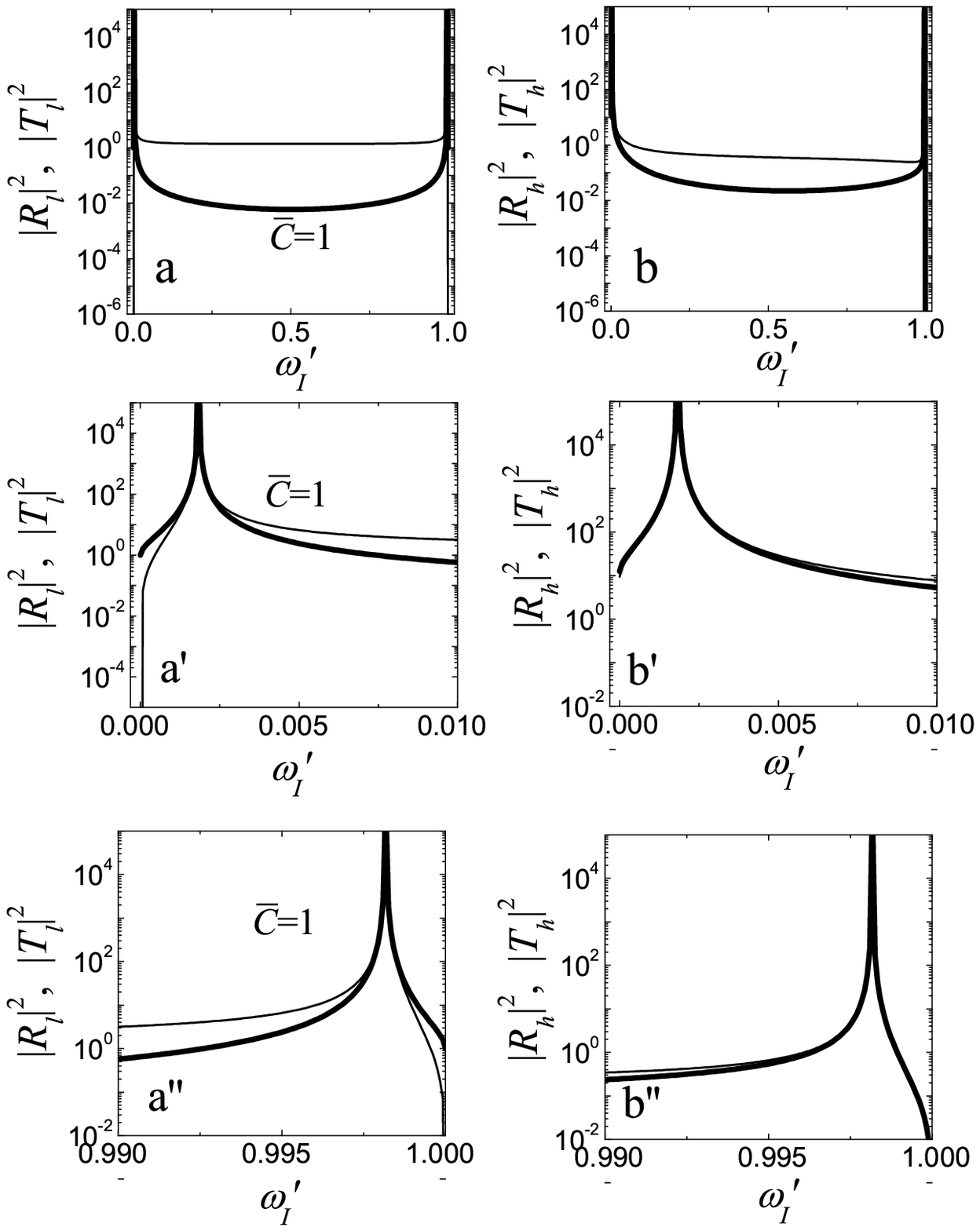}
\caption{Same as in Fig.~\protect\ref{Fig5}, but for $\overline{C}=1$.}
\label{Fig6}
\end{figure}

For $\overline{C}=1$, the reflection and transmission coefficients are
presented in Fig.~\ref{Fig6}. All four coefficients diverge at $\omega
_{I}^{\prime }=0.0022$ ($\omega _{I}=0.88$ ) and $\omega _{I}^{\prime
}=0.999 $ ($\omega _{I}=7.1225$) because at these two points the denominator
$D$ in Eq.~(\ref{coeffwave}), which is common for all coefficients,
vanishes. On (a'),(b') we show the blowups of (a),(b) for the maxima close
to $\omega _{I}^{\prime }=0$, while in (a''),(b'') for the maxima close to $%
\omega _{I}^{\prime }=1$. The coefficient $T_{l}$ drops to zero at $\omega
_{I}^{\prime }=0$ ($\omega _{I}=0.866$).

\begin{figure}[tbp]
\includegraphics[width=\columnwidth]{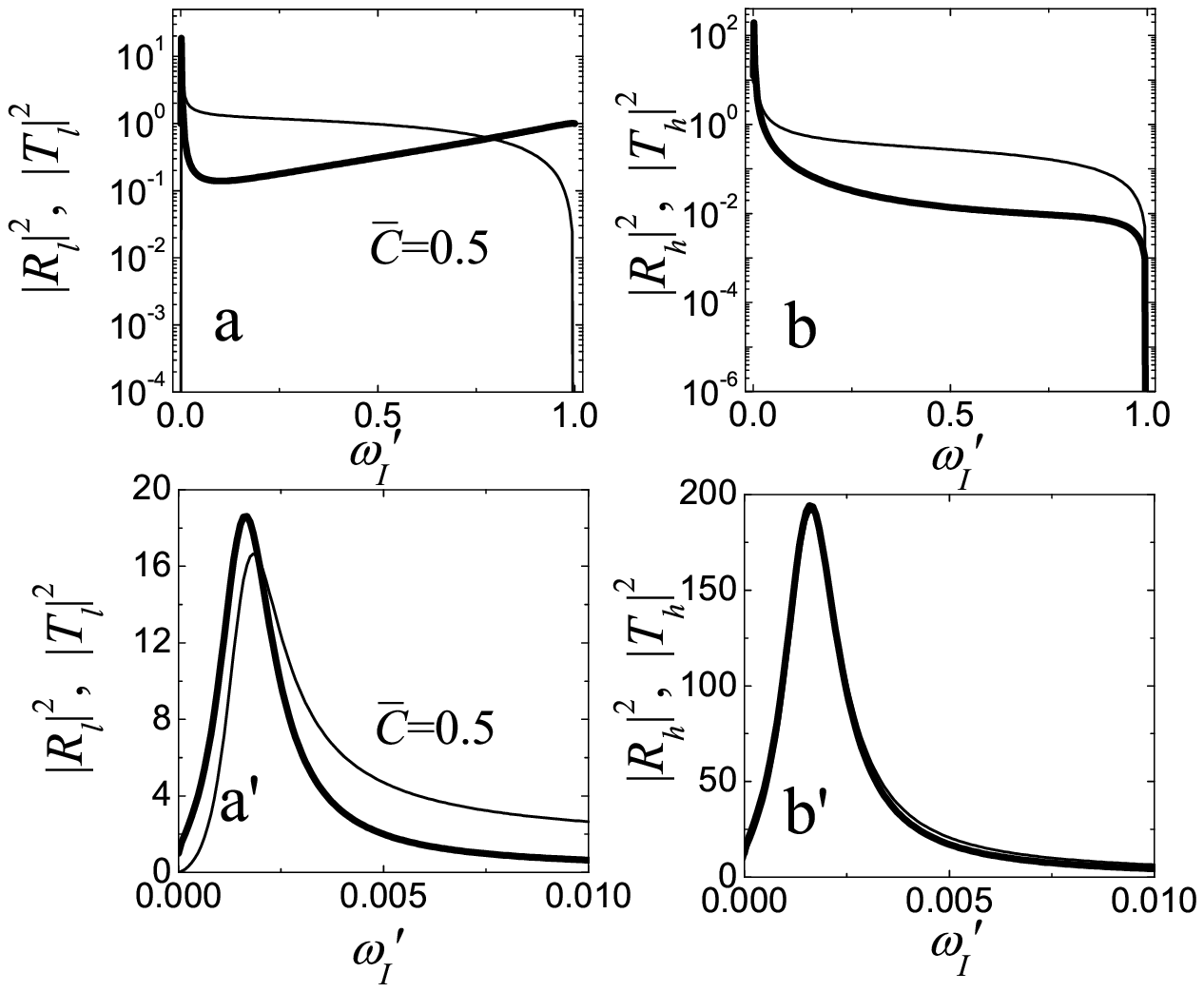}
\caption{Same as in Figs. \protect\ref{Fig5} and \protect\ref{Fig6}, but for
$\overline{C}=0.5$.}
\label{Fig7}
\end{figure}

For $\overline{C}=0.5$, the reflection and transmission coefficients are
displayed in Fig.~\ref{Fig7}. All the four coefficients have finite maxima
at $\omega _{I}^{\prime }=0.00446$ ($\omega _{I}=0.894$), and $T_{l}$
vanishes at $\omega _{I}^{\prime }=0$ ($\omega _{I}=0.86603$).

The above results can be summarized by saying that for $\overline{C}\neq 1$
all the four reflection and transmission coefficients attain finite maxima
at $\omega _{I}^{\prime }$ close (but not equal) to zero. For the special
case of $\overline{C}=1$ all four coefficients diverge at $\omega
_{I}^{\prime }$ close to zero and close to unity.

It is relevant to note that, if $R_{l}=0$ at a particular value of $\omega
_{I}$, then $T_{l}=1$, and, conversely, $R_{l}=1$ if $T_{l}=0$. These cases
correspond to the full transmission and full reflection for the LF waves,
respectively. However, at $\gamma \neq 0$ the HF reflection and transmission
waves are excited too, therefore, in fact, these cases imply not full
transmission and reflection, but rather full conversion of LF waves into
their HF counterparts.

\begin{figure}[tbp]
\includegraphics[width=\columnwidth]{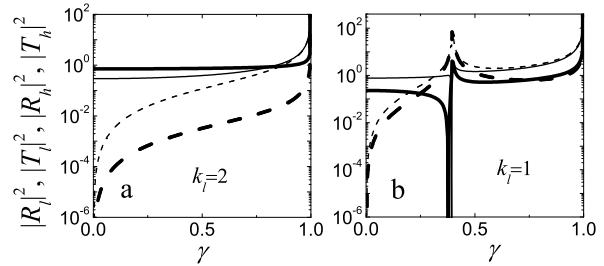}
\caption{$|R_{l}|^{2}$ (thick line), $|T_{l}|^{2}$ (thin line), $|R_{h}|^{2}$
(dashed thick line), $|T_{h}|^{2}$ (dashed thin line) as functions of $%
\protect\gamma $ in the case of low-frequency incident CW for $C_{1}=2$, $%
C_{2}=1$. (a) $k_{l}=2$ with real $k_{h}$, and (b) $k_{l}=1$ with $k_{h}$
changing from imaginary to real at $\protect\gamma =0.39$.}
\label{Fig8}
\end{figure}

It is also interesting to consider the effect of amplification of the
reflected and transmitted waves, which is possible in the presence of the
gain and loss, $\gamma >0$ \cite{Miroshnichenko}. To this end, we fix $%
C_{1}=2$ and $C_{2}=1$ and vary the gain/loss parameter within the range of $%
0\leq \gamma <1$ at fixed $k_{l}$, which leads to the variation of $\omega
_{I}$ from $4C_{1}(1-\cos {k_{l}})-1$ to $4C_{1}(1-\cos {k_{l}})$.
Accordingly, $k_{h}$ changes too, taking both real and imaginary values. For
this case, coefficients $|R_{l}|^{2}$, $|T_{l}|^{2}$, $|R_{h}|^{2}$, and $%
|T_{h}|^{2}$ are shown versus $\gamma $ in Fig.~\ref{Fig8}, for $k_{l}=2$
(a) and $k_{l}=1$ (b). Note that $k_{h}$ remains real in (a) for all $\gamma
$, while in (b) $k_{h}$ changes from imaginary to real at $\gamma =0.39$. As
seen in the figure, in (a) all the reflection and transmission coefficients
increase with $\gamma $. On the other hand, in panel (b) they feature
additional extrema in a vicinity of the point where $k_{h}$ changes from
imaginary to real.

\begin{figure}[tbp]
\includegraphics[width=\columnwidth]{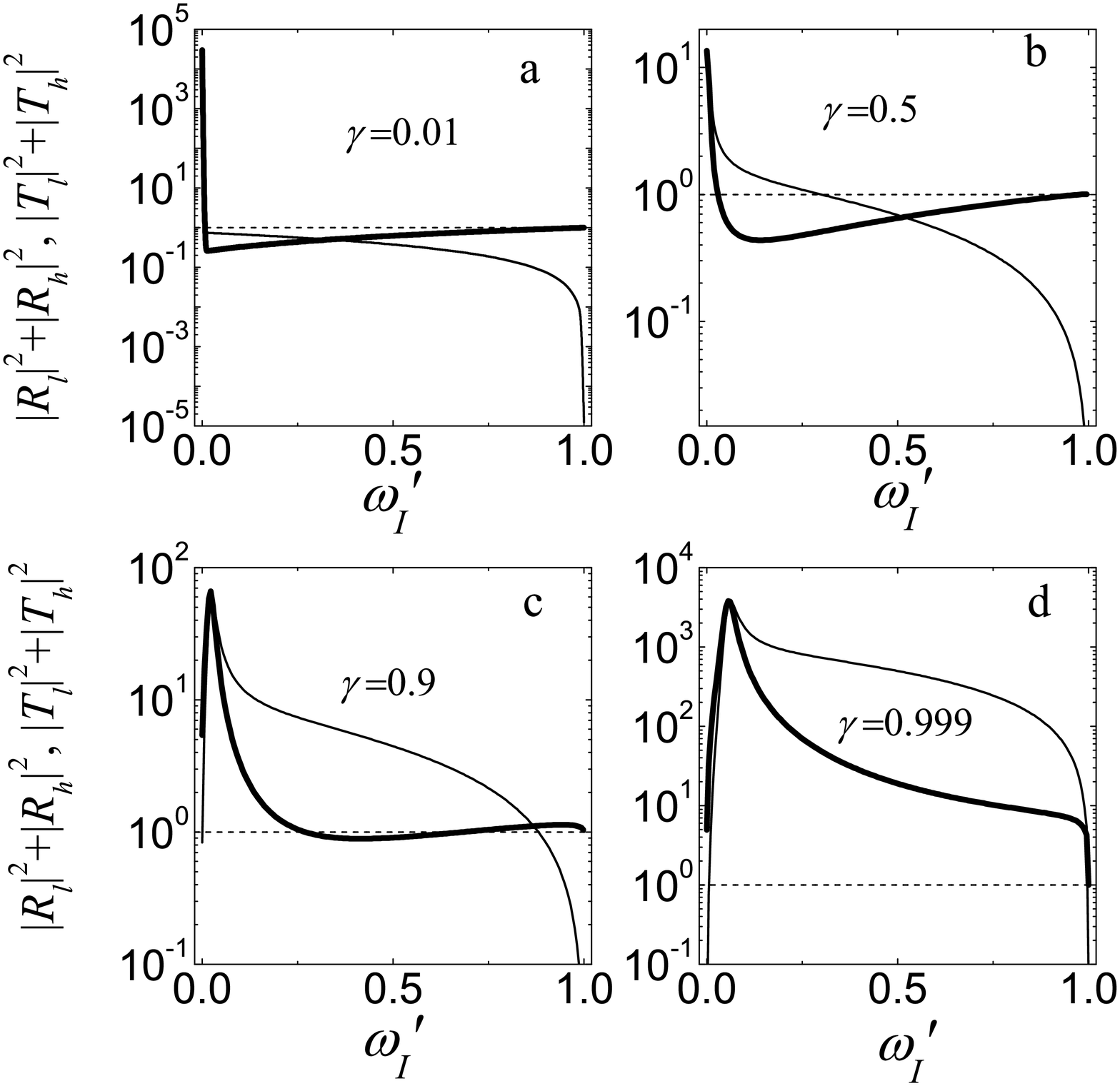}
\caption{$|R_{l}|^{2}+|R_{h}|^{2}$ and $|T_{l}|^{2}+|T_{h}|^{2}$ (thick and
thin lines, respectively) as functions of $\protect\omega _{I}^{\prime }$ in
the case of low-frequency incident CW for $C_{1}=2$, $C_{2}=1$, and $\protect%
\gamma =0.01$ (a), $\protect\gamma =0.5$ (b), $\protect%
\gamma =0.9$ (c), $\protect\gamma =0.999$ (d) Note that the vertical scale
is logarithmic.}
\label{Fig9}
\end{figure}

The amplification of the reflected and transmitted waves can be defined in
terms of their total intensities. Thus, in the case of the incident LF CW,
when both reflected and transmitted HF waves are of CW type, the
amplification of reflection and transmission takes place when $%
|R_{l}|^{2}+|R_{h}|^{2}>1$ and $|T_{l}|^{2}+|T_{h}|^{2}>1$, respectively (as
shown in Fig.~\ref{Fig9}). We notice that the range of the incident-wave's
frequency, $\omega _{I}$, where the amplification occurs, increases with $%
\gamma $. Also note that in Fig.~\ref{Fig9} (a) the reflection and
transmission coefficients feature sharp maxima in a vicinity of
$\omega_{I}^{\prime}=0$. In the limit $\gamma=0$, as it has been
already mentioned, the system becomes conservative and these
maxima disappear.

\section{Conclusions}

\label{Conclusions}

We have studied the scattering of linear waves on a domain wall
introduced into a waveguide array composed of
$\mathcal{PT}$-symmetric waveguide pairs. Such arrays support the
propagation of HF\ (high-frequency) and LF\ (low-frequency) waves.
Considering incident LF waves of various types (continuous waves
or unstaggered and staggered exponential waves), we have derived
the corresponding reflection and transmission coefficients and
analyzed their dependence on the system parameters. The case of
the HF incident wave can be analyzed similarly.

We have found that the LF incident wave generates both LF and HF reflected
and transmitted waves, provided that the gain and loss are present ($\gamma
>0$). We also demonstrated that both reflected and transmitted waves can be
substantially amplified, provided that the gain $\gamma >0$ is present. The
range of the incident-wave's frequency where the amplification takes place
expands with the increase of $\gamma$.

Our results suggest that the use of $\mathcal{PT}$-symmetric elements in
waveguide arrays offers various possibilities for manipulations of optical
signals in photonic lattices. It may be interesting to add nonlinearity to
the system. In addition to the formation of solitons \cite{we}, the
nonlinearity may give rise to spontaneous symmetry breaking \cite%
{Miroshnichenko}. Obviously, these effects may strongly affect the
scattering problem. Finally, a challenging problem is to extend the analysis
to the case of two-dimensional $\mathcal{PT}$-invariant networks.

\section*{Acknowledgments}

Sergey V. Suchkov thanks Liya Z. Khadeeva for the discussions. S.~V.~Suchkov
and S.~V.~Dmitriev acknowledge financial support from the Russian Foundation
for Basic Research, grant 11-08-97057-p-povolzhie-a. The work was partially
supported by the Australian Research Council.

\end{document}